\documentclass[10pt,a4paper,twocolumn]{article}
\usepackage{lmodern}

\usepackage{amssymb}
\usepackage{lipsum}
\usepackage{xcolor}
\usepackage{mathrsfs}
\usepackage{amsmath}

\usepackage{abstract} % 

\usepackage{graphicx}  %% 

\usepackage[small]{caption}

\author{Filippo Belloni$^{1}$, Mauro Chiesa$^{1,*}$, Clara Del Pio$^{2}$, Fulvio Piccinini$^{1}$}

\date{%
  \small{$^1$INFN Sezione di Pavia, Via Bassi 6, Pavia, 27100, Italy\\
  $^2$Physics Division, Brookhaven National Laboratory, Upton, 11973, NY, USA\\[2ex]%
  $^*$corresponding author: {\tt mauro.chiesa@pv.infn.it }}
}

\title{Photon induced corrections to Drell-Yan production in POWHEG}

\begin{document}

\twocolumn[
\maketitle % 
\begin{onecolabstract}
High-precision theoretical predictions for both signal and background processes are crucial for the precision physics program at the LHC and HL-LHC.
A prime example is the determination of the effective leptonic weak mixing angle from neutral current Drell-Yan via template-fit methods, 
where Monte Carlo templates directly serve as input to the measurement, making theoretical uncertainties part of the systematic error affecting the experimental results.
In this paper, we focus on photon-induced corrections to neutral current Drell-Yan, as well as on the irreducible $\gamma\gamma \to l^+l^-$ background.
We present the implementation of the photon-initiated contributions in the {\tt Z\_ew-BMNNPV} Monte Carlo generator for Drell-Yan production at NLO QCD + NLO EW accuracy 
matched to QED and QCD parton showers within the POWHEG framework. We discuss phenomenological results for the dilepton invariant-mass distribution and the forward-backward asymmetry
in the $Z$-resonance region and in the TeV range.
\\
\\
\\
  {\bf Keywords:} Electroweak corrections, photons, Monte Carlo generators, NLO+parton shower, Drell-Yan, weak-mixing angle.
\\
\\
\end{onecolabstract}
]

\section{Introduction}
\label{introduction}

The precise determination of the cross section and differential distributions of the  Drell-Yan (DY) processes is one of the goals of the LHC and future hadron colliders. 
DY measurements serve multiple purposes: they enable the precise determination of key Standard Model parameters $-$ such as the W-boson mass ~\cite{CDF:2013dpa,D0:2013jba,CDF:2022hxs,LHCb:2021bjt,LHC-TeVMWWorkingGroup:2023zkn,ATLAS:2024erm,CMS:2024lrd} and the sine of the leptonic effective weak mixing angle~\cite{CDF:2018cnj,ATLAS:2015ihy,LHCb:2024ygc,CMS:2024ony} $-$ and they provide an essential contribution for PDF determination.
Furthermore, DY processes are very important in the context of New Physics searches, ranging from the indirect searches as a by-product of the precision physics program at the weak scale, to the direct searches in the high dilepton invariant-mass region. 

The precision physics program based on DY measurements demands extremely precise theoretical predictions for $W$ and $Z$ boson production. 
In particular, measurements based on template-fit methods 
$-$ like the determination of the weak mixing angle $-$ 
rely on theory predictions via the Monte Carlo templates. 
It is thus crucial to have theory uncertainties under control, 
as they are part of the overall systematic error 
and may ultimately become a limiting factor 
on the achievable experimental accuracy.

This context has motivated a major effort within the theory community, pushing the frontier of the predictions for DY to N$^2$LO~\cite{Hamberg:1990np,Anastasiou:2003ds,Melnikov:2006kv,Melnikov:2006di,Catani:2009sm,Gavin:2010az,Gavin:2012sy} and N$^3$LO in QCD~\cite{Duhr:2020sdp,Duhr:2020seh,Duhr:2021vwj,Baglio:2022wzu,Chen:2021vtu,Chen:2022lwc,Chen:2022cgv,Neumann:2022lft}. 
Significant progress has also been achieved in the treatment of 
electroweak (EW) effects, including  two-loop mixed QCD-EW  corrections~\cite{Delto:2019ewv, Cieri:2020ikq,
  Bonciani:2016wya, Bonciani:2019nuy, Bonciani:2020tvf, Bonciani:2021iis,
  Buccioni:2020cfi, Behring:2020cqi, Dittmaier:2014qza, Dittmaier:2015rxo,
  Dittmaier:2020vra, Dittmaier:2024row, deFlorian:2018wcj, Bonciani:2021zzf,
  Buccioni:2022kgy, Armadillo:2022bgm, Buonocore:2021rxx, Armadillo:2024nwk,
  Behring:2021adr} and 
 two-loop electroweak corrections in pole approximation~\cite{Chen:2022dow,Freitas:2025vax}.  
The first results beyond pole approximation have been presented in~\cite{Armadillo:2025mfx,Armadillo:2026wdb}. 
Combined QCD and EW resummation strategies have also been investigated~\cite{Cieri:2018sfk,Billis:2019evv,Autieri:2023xme,Buonocore:2024xmy,Autieri:2025tho}. 
N$^2$LO QCD corrections have been combined with
NLO EW effects in \cite{Li:2012wna,Grazzini:2017mhc, Buonocore:2019puv, Grazzini:2019jkl}.
Several algorithms have been developed to match NNLO QCD calculations to parton showers (PS)~\cite{Monni:2019whf, Monni:2020nks,Hamilton:2013fea,
  Alioli:2013hqa, Hoche:2014uhw, Hoche:2014dla}
and there are ongoing efforts to extend these matching strategies to electroweak corrections~\cite{Belloni:2026vnv}. 
NLO QCD corrections matched to PS can be computed with many different tools~\cite{Nason:2004rx, Frixione:2007vw, Alioli:2010xd, Jezo:2015aia,
  Frixione:2002ik, Alwall:2014hca, Bellm:2015jjp, Platzer:2011bc,
  Hoeche:2011fd, Sherpa:2019gpd, Bauer:2008qh, Alioli:2013hqa}. 
NLO EW corrections have been known for a long
time~\cite{Dittmaier:2001ay, Baur:2004ig, Zykunov:2006yb, Arbuzov:2005dd,
  CarloniCalame:2006zq, Baur:2001ze, Zykunov:2005tc, CarloniCalame:2007cd,
  Arbuzov:2007db, Dittmaier:2009cr, Baur:1998kt,Arbuzov:2023afc} and are available in both
dedicated Monte Carlo codes~\cite{CarloniCalame:2006zq, CarloniCalame:2007cd,
  Baur:1997wa, Baur:2001ze, Arbuzov:2007db, Andonov:2008ga, Bardin:2012jk,
  Bondarenko:2013nu, Arbuzov:2015yja, Arbuzov:2016wfy, Arbuzov:2022sep,
  Dittmaier:2009cr, Campbell:2016dks,Alioli:2016fum} and automated
tools~\cite{Alwall:2014hca, Frederix:2018nkq, Biedermann:2017yoi,
  Sherpa:2024mfk,Denner:2026phn}. 
  Multiple
photon radiation effects have been studied in~\cite{Placzek:2003zg,
  CarloniCalame:2003ux, CarloniCalame:2007cd, Brensing:2007qm,
  Dittmaier:2009cr}. 

The {\tt Z\_ew-BMNNPV} package~\cite{Barze:2013fru,Chiesa:2019nqb,Chiesa:2024qzd} of the {\tt POWHEG-BOX-V2}~\cite{Nason:2004rx,Frixione:2007vw,Alioli:2010xd} framework is a Monte Carlo event generator for neutral-current Drell-Yan (NC DY) at NLO QCD$+$NLO EW (plus universal fermionic contributions beyond $\mathcal{O}(\alpha)$ as documented in~\cite{Chiesa:2024qzd}) matched to QCD and QED PS\footnote{Independent implementations of the NLO QCD$+$NLO EW corrections to DY can be found in~\cite{Bernaciak:2012hj,Muck:2016pko}.}.
The generator provides a resonance-aware version of the matching algorithm~\cite{CarloniCalame:2016ouw} based on the one developed in~\cite{Jezo:2015aia}.
Very recently, the matching of the NLO EW corrections to PS within the MC@NLO framework was discussed in~\cite{Flower:2026lui}.

The $\gamma\gamma\to l^+l^-$ process is an irreducible background to NC DY production at hadron colliders. 
Compared to the $q\bar{q}$ channel, $\gamma\gamma$-initiated processes are heavily suppressed due to the photon PDF compared to the quark PDFs. 
For this reason, these contributions are typically simulated at LO ($+$PS) and subtracted in the experimental analyses.  
However, when considering NC DY beyond the LO in the EW coupling, the separation between the $q\bar{q}$ and the $\gamma\gamma$ channels is less straightforward.  
In fact, two classes of processes contribute to the real $\mathcal{O}(\alpha)$ corrections to the $q\bar{q}$-initiated channel: 
photon bremsstrahlung ($q\bar{q} \to l^+l^- \gamma$)  and  $q\gamma$ initiated processes ($q\gamma \to l^+l^-q$, where the final-state parton is integrated over the full phase space).
By examining the collinear limit of the latter, one can see that the $q\gamma$-initiated processes can be interpreted as real corrections to both the $q\bar{q}$ and $\gamma\gamma$ channels.

In this paper, we present the implementation of the LO ($\gamma\gamma$) and NLO ($q\gamma$) $\gamma$-initiated contributions in the {\tt Z\_ew-BMNNPV} Monte Carlo event generator in the POWHEG framework.
While the numerical impact of the $\gamma\gamma$ and $q\gamma$ channels has been studied in the literature~\cite{Dittmaier:2009cr,Harland-Lang:2016kog,Accomando:2016tah,Accomando:2016ehi,Harland-Lang:2016apc,Harland-Lang:2020veo,Zykunov:2024gma}, 
in this paper we present the first implementation of $\gamma$-induced contributions to Drell-Yan at NLO matched to PS in POWHEG\footnote{Photon and lepton induced processes in POWHEG have been previously studied in~\cite{Buonocore:2021bsf} in the context of BSM searches.}.
In Section~\ref{sect:implementation}, we describe the modifications to the POWHEG routines required to generate $\gamma$-initiated events and in particular to deal with the initial-state singular region related to the $q\to q\gamma$ splitting where the quark is the radiated parton.
In Section~\ref{sect:nums} we validate the implementation and show results for the dilepton invariant-mass and forward-backward asymmetry distributions in the $Z$ resonance region and in the high invariant mass regime. 

\section{Implementation of the $\gamma$-induced  processes}
\label{sect:implementation}
Real  radiation processes of the form $q\gamma \to l^+l^-q$ (where $q$ generically denotes both quarks and antiquarks)
enter at the same perturbative order as photon bremsstrahlung, 
but they are typically suppressed by the photon PDF compared to quark PDFs. The 
$q\gamma \to l^+l^-q$ processes develop initial state (IS) singularities 
from $q \to q \gamma$ and $\gamma \to q \bar{q}$ splittings as the final state quark 
becomes collinear with the beams. 
While finite lepton masses prevent singular $\gamma \to l^+l^-$ splittings, 
they nonetheless give rise to log-enhanced $t$-channel contributions.  
IS singularities in $q\gamma \to l^+l^-q$ processes are  purely collinear: 
they have no virtual counterpart and are reabsorbed in PDF collinear remnants.  

The implementation of the  $q\gamma \to l^+l^-q$  processes in 
the {\tt Z\_ew-BMNNPV} package of {\tt POWHEG} requires the corresponding matrix elements 
(obtained from the bremsstrahlung ones by crossing) and the definition of a new underlying Born (UB) 
process, namely $\gamma\gamma\to l^+l^-$, as the collinear $q\to q \gamma$ splittings can be mapped 
on both quark and $\gamma$ initiated dilepton production. This means that both processes are required to build the FKS~\cite{Frixione:1995ms,Frixione:1997np} subtraction terms. 
All the required modifications to the PDF collinear remnant terms have been implemented 
in a local version of the {\tt sigcollremn.f} file. 
Though lepton masses prevent $t$-channel singularities in the $\gamma\gamma\to l^+l^-$ UB, 
the numerical values of the electron and muon masses (compared to typical large values of the kinematical invariants)
may lead to numerical instabilities, 
that have been addressed by means of a combination of strategies, spanning from 
a dedicated importance sampling on the lepton azimuthal angle, to \emph{ad-hoc} suppression factors 
and generation cuts in the very forward rapidity lepton region in the $2 \to 2$ phase space. 
The latter strategy turns out to be the most effective for numerical stability, 
but forces us to consider only predictions in a reasonably central region of the DY phase-space.

The generation of events in the POWHEG framework requires further modifications 
to deal with the IS $q \to q \gamma$ and $\gamma \to q \bar{q}$ splittings. 
In fact, the default upper bounding function for initial-state radiation (ISR) reads:
\begin{equation}
U(x,y)=N\frac{\alpha_S(k_T^2)}{(1-x)(1-y^2)},
\label{eq:ISRUPB}
\end{equation}
with $k_T^2=\frac{s_b}{4x} (1-x)^2(1-y^2)$,  
$s_b$ being the Born center of mass energy squared, $x=1-\xi$, 
while $\xi$ and $y$ are the radiation variables in the notation of~\cite{Alioli:2010xd}. 
Eq.~(\ref{eq:ISRUPB}) represents an upper bound for the ratio 
\begin{equation}
\frac{R_{ij}(\Phi_{UB},\Phi_{rad})}{B_{kl}(\Phi_{UB}) } \, ,
\end{equation}
where $\Phi$ stands for the set of momenta of the considered phase-space point.
We use the notation $X_{ij}=|{\cal X}_{ij}|^2 \otimes p_i \otimes p_j$, where
${\cal X}_{ij}$ is the matrix element for the Born or real process with IS of flavour $i$ and $j$, 
and $p_{i(j)}$ are the corresponding PDFs. 

The upper bound in~(\ref{eq:ISRUPB}) has been obtained from the soft/collinear limit of singular splittings 
and from an estimate of the $(x,\mu_F)$ dependence of the gluon and quark PDFs. 
In our calculation, we need the IS upper bounding functions for the ratios $R_{q\gamma}/B_{qq}$ and  
$R_{q\gamma}/B_{\gamma\gamma}$ besides the $R_{qq}/B_{qq}$ already present in {\tt Z\_ew-BMNNPV} 
and discussed in~\cite{Barze:2012tt}.
Although the form of the singular $\gamma-q$ splittings can be taken from the $g-q$ ones 
via abelianization and replacing $\alpha_S(k_T^2) \to \alpha_0$, the bounds of the PDF ratios 
involving  the photon cannot be taken directly from the ones obtained in the case of QCD partons. 
For the ratios involving the $B_{qq}$ UB one can still use eq.~(\ref{eq:ISRUPB}), as the photon PDF 
is smaller than the gluon one. 
However, for the singular regions involving the $B_{\gamma\gamma}$ UB we need to bound the ratio $p_q/p_{\gamma}$, which is much larger than the corresponding ratio in the QCD case, namely $p_q/p_g$. 
A simple, though not very elegant, strategy could be to study the numerical value of the photon PDF 
compared to the gluon and quark ones for a suitable range of $x$ and $\mu_F$ values 
and find a value $k_\gamma$ such that $k_\gamma p_\gamma (x,\mu_F) > p_{q/g} (x,\mu_F)$. 
In practice, for $x > 10^{-6}$ and $\mu_F$ in the range between $\mathcal{O}(1~{\rm GeV})$ and $\mathcal{O}(10~{\rm TeV})$, 
$k_\gamma$ can be safely set to about 500.  
In this way, the upper bounding function for ISR involving the $B_{\gamma\gamma}$ UB is the usual one with the replacement $\alpha_S \to k_\gamma \alpha_0$.  
The factor $k_\gamma$ is consistently taken into account in the other stages of event generation, namely in subroutines 
{\tt inc\_norms} and {\tt gen\_rad\_isr}.  For the latter, we also derived the expression of the $p_T$ integral of the upper bounding function with constant coupling $\alpha_0$ in the case of the $\gamma\gamma$ UB. 
We stress that the strategy outlined above, though not very efficient or completely general, 
still works reasonably well for DY mainly because of the relative size of the $\gamma\gamma\to l^+l^-$ UB 
compared to the $qq\to l^+l^-$ one in the phase-space region of interest. 
As a last remark, even though the $qq$ and $\gamma\gamma$ UB processes contribute to dilepton production at the same order in $\alpha$, 
the contribution of the latter process is similar in size to the NLO EW corrections to the $qq$ channel 
and for this reason the $\gamma\gamma$ process is usually considered at LO. 
However, in the code, we added the possibility to compute the NLO photonic corrections to $\gamma\gamma \to l^+l^-$ 
by implementing the real and virtual~\cite{Denner:1998tb} matrix elements. 
For this process, only FSR radiation off massive leptons is present and it can be treated as in the case of the $qq$ underlying Born.

\section{Validation and numerical results}
\label{sect:nums}

In this section, we provide results for di-muon production at the LHC at $13$~TeV focusing on the impact of $\gamma$-initiated processes.
In the simulations, muons are massive, with $m_\mu=0.105$~GeV, and only bare muons are considered in the analyses.
The calculation is performed in the $(G_\mu,M_W,M_Z)$ scheme, with the pole $W$ and $Z$ masses and widths 
derived from the corresponding on-shell values $M_Z^{OS}=91.1876$~GeV, $\Gamma_Z^{OS}=2.4952$~GeV, 
$M_W^{OS}=80.385$~GeV, $\Gamma_W^{OS}=2.085$~GeV. 
The complex-mass scheme~\cite{Denner:1999gp,Denner:2005fg,Denner:2006ic} is used to deal with 
the presence of the unstable $Z$ boson in the $q\bar{q}$ channel.
If not stated otherwise, the {\tt NNPDF31\_nlo\_as\_0118\_luxqed}~\cite{Manohar:2016nzj,Manohar:2017eqh,Bertone:2017bme} PDF set\footnote{More recent 
implementations of the photon PDF can be found, for instance, in~\cite{Xie:2021equ,Cridge:2021pxm,NNPDF:2024djq}} 
is used via the {\tt LHAPDF} library~\cite{Buckley:2014ana} with factorization scale $\mu_F=M_Z$ and 
the calculation is performed in the $\overline{\rm MS}$ factorization scheme.

We label NLO and NLO+PS the results obtained at fixed-order
and at NLO matched to Parton Shower accuracy in the POWHEG framework, respectively. 
In particular, $q\bar{q}_{\rm NLO}$ stands for NLO QCD + NLO EW corrections (plus the universal fermionic higher-order effects detailed in~\cite{Chiesa:2024qzd}) in the $q\bar{q}$ channel, 
$q\bar{q}_{\rm NLO}$ + $\gamma\gamma$ also includes the $\gamma\gamma$ contribution at LO, 
while $q\bar{q}_{\rm NLO}$ + $\gamma\gamma$ + $q\gamma$ corresponds to the calculation for $q\bar{q}_{\rm NLO}$ 
plus the contribution of the real $q\gamma$ processes and the $\gamma\gamma$ UB 
needed to build the IR subtraction. 
The curves labeled $q\bar{q}_{\rm NLO}$ + $\gamma\gamma_{\rm QED}$ also include the 
NLO QED corrections to the $\gamma \gamma$ channel; while these corrections 
have been investigated in the literature~\cite{Bohm:1986dn,Denner:1993ts,Dittmaier:1993bj,Denner:1995ar,Denner:1998tb,Zykunov:2024gma}, we include them in the plots 
to assess their relative size compared to the leading $q\gamma$-initiated contributions.
The renormalization scale for the QCD corrections is set to the $Z$ mass. 
Virtual photonic contributions and real $\mathcal{O}(\alpha)$ radiation are computed with 
two powers of $\alpha$ derived from $G_\mu$ and one set to $\alpha_0$. 
The values of the parameters needed for the weak corrections are taken from~\cite{Chiesa:2024qzd}.
At the event level, {\tt POWHEG} generates the hardest ISR and FSR radiations, 
while {\tt PYTHIA}~\cite{Sjostrand:2006za,Sjostrand:2014zea} is used to generate QCD and QED emissions softer than the POWHEG ones. 
The {\tt Z\_ew-BMNNPV} package also provides an interface to {\tt PHOTOS}~\cite{Golonka:2005pn,Barberio:1990ms,Barberio:1993qi} for the QED FSR: 
when matching is performed, 
the residual differences between simulations based on the PHOTOS and PYTHIA showers are small, as documented in~\cite{CarloniCalame:2016ouw}.
The event selection is taken from~\cite{Dittmaier:2024row} and requires $p_T^{\mu}>25$~GeV, 
$|y^{\,\mu}|<2.5$, 
and $M_{\mu^+\mu^-}>50$~GeV. 
We mainly focus on the dilepton invariant-mass distribution and on the forward-backward 
asymmetry defined as:
\begin{eqnarray}
  A_{FB}(M_{ll})&=&\frac{d\sigma^F/dM_{ll} - d\sigma^B/dM_{ll}}
 {d\sigma^F/dM_{ll} + d\sigma^B/dM_{ll}}, 
  \label{eq:afb} \\
  d\sigma^F/dM_{ll} &=& \int_0^1 dc\,  \frac{d\sigma}{dM_{ll}\, dc}\, ,
  \label{eq:sigma_F} \\
  d\sigma^B/dM_{ll} &=& \int_{-1}^0 dc\, \frac{d\sigma}{dM_{ll}\, dc}\, ,
  \label{eq:sigma_B}
\end{eqnarray}
where $c$ is the cosine of the lepton scattering angle in the Collins-Soper
frame. We present the results in two invariant-mass regions: the first corresponding to the $Z$-peak region 
and the second covering the high invariant-mass tail. 
Although the asymmetry is usually studied in the peak region and used for the 
extraction of the leptonic effective weak mixing angle, at high invariant-masses 
it can provide information on the running of the mixing angle~\cite{Amoroso:2023uux} 
as well as on possible New Physics~\cite{Accomando:2016ehi}. 

\begin{figure}
	\includegraphics[width=0.47\textwidth]{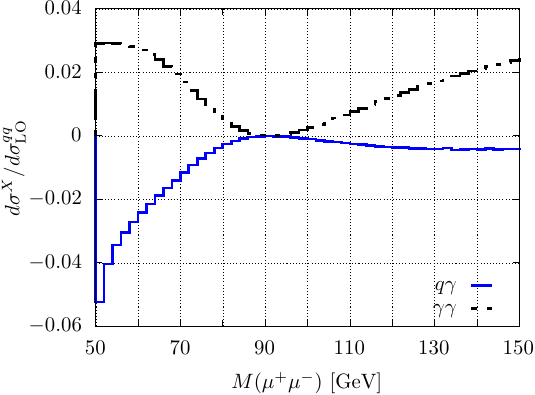} \\
    \includegraphics[width=0.47\textwidth]{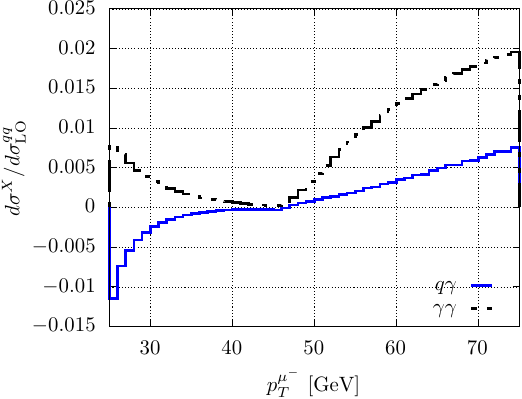}	
	\caption{Relative impact of $q\gamma$ and $\gamma\gamma$ 
    contributions w.r.t. the LO $q\bar{q}$ differential cross section as a function of $M(\mu^+\mu^-)$ and $p_T^\mu$ in the setup of~\cite{Dittmaier:2009cr}.} 
	\label{fig:DH}%
\end{figure}

We consider the setup of~\cite{Dittmaier:2009cr} for the validation of the implementation of the real $q\gamma$ 
and $\gamma\gamma$ UB processes. 
Figure~\ref{fig:DH} shows the relative impact of the photon-induced processes with respect to the 
$q\bar{q}$ channel computed at LO as a function of the muon-pair invariant mass and muon transverse momentum. The plots should be compared to Figs.~12 and~13 of~\cite{Dittmaier:2009cr}. 
Following~\cite{Dittmaier:2009cr}, the results in Fig.~\ref{fig:DH} are obtained employing the {\tt MRST2004qed}~\cite{Martin:2004dh} PDF set and the DIS factorization scheme. 
Moreover, the LO $\gamma\gamma$ curve is computed by setting the photon-fermion coupling to $\alpha_0$. 
As further validation, we compared the predictions of the {\tt Z\_ew-BMNNPV} package 
with the results of an independent Monte Carlo code (derived from the one used in~\cite{Chiesa:2018lcs,Chiesa:2020ttl}) 
based on {\tt Recola}~\cite{Actis:2012qn,Actis:2016mpe} 
for the calculation of the matrix elements and on the Catani-Seymour dipoles~\cite{Catani:1996vz} 
for the subtraction of infrared singularities.

\begin{figure}
	\includegraphics[width=0.47\textwidth]{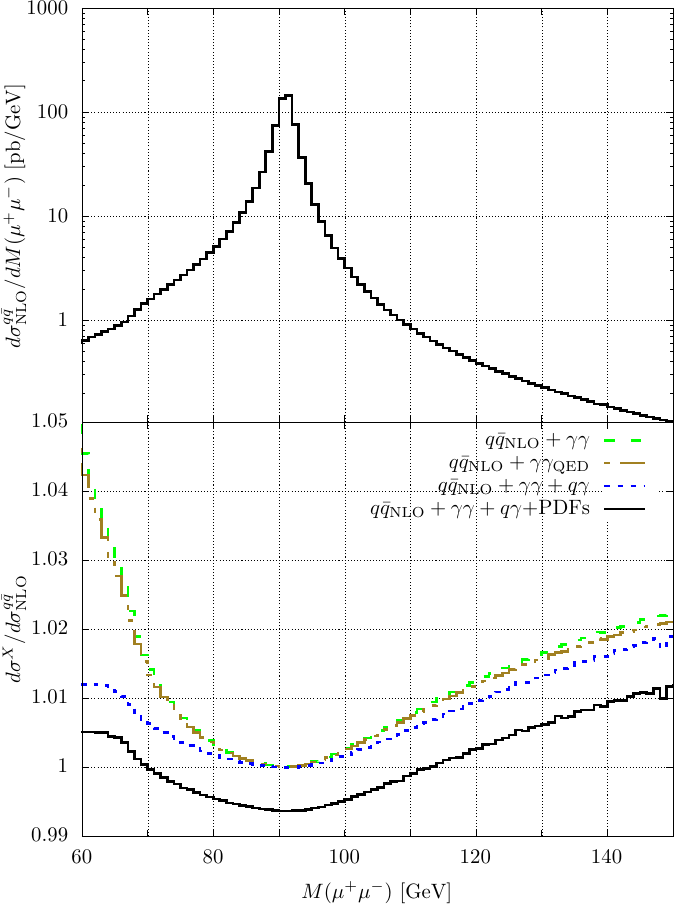} 
	\caption{Ratios of predictions for the invariant mass distribution with (numerator) and without (denominator) photon-induced processes. The $\gamma$-initiated processes entering the dashed green, the dot-dashed red, and the dotted blue lines are the $\gamma\gamma$ underlying Born, the $\gamma\gamma$ process at NLO QED, and the real $q\gamma$ corrections plus the corresponding $\gamma\gamma$ UB, respectively. For all the curves, the numerators include the $q\bar{q}$ channel at NLO QCD+NLO EW plus universal fermionic $\mathcal{O}(\alpha^2)$ corrections computed using the  {\tt NNPDF31\_nlo\_as\_0118\_luxqed} PDF set. 
    The denominators contain only the $q\bar{q}$ channel at NLO QCD+NLO EW accuracy plus universal fermionic corrections. For the dashed green, the dot-dashed red, and the dotted blue curves, the {\tt NNPDF31\_nlo\_as\_0118\_luxqed} PDF set is also used in the denominator, 
    while the solid black line has the same content as the dotted blue one, with the only difference that the photon contribution to the PDF evolution is not included in the denominator, which is computed using the {\tt NNPDF31\_nlo\_as\_0118} PDF set. Upper panel: $M(\mu^+\mu^-)$ distribution computed at NLO QCD+NLO EW plus fermionic higher orders without $\gamma$-induced processes.} 
	\label{fig:nlomll}%
\end{figure}

\begin{figure}
	\includegraphics[width=0.47\textwidth]{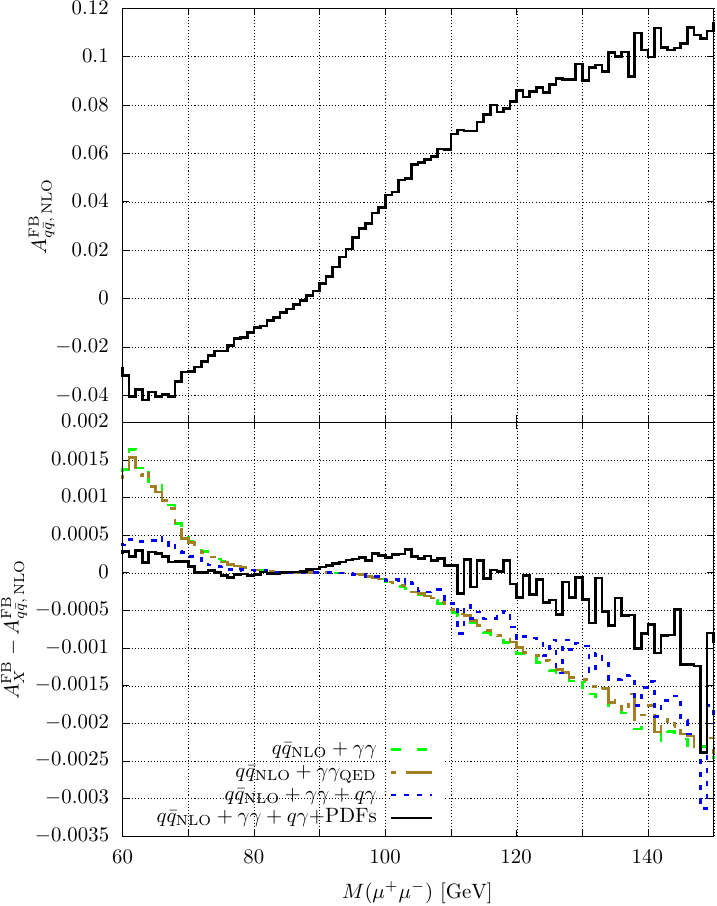} 
	\caption{Differences of predictions for the forward-backward asymmetry distribution as a function of $M(\mu^+\mu^-)$ with and without photon-induced processes. Same conventions as in Fig.~\ref{fig:nlomll}. 
    Upper panel: $A_{\rm FB}$ distribution computed at NLO QCD+NLO EW plus fermionic higher orders without $\gamma$-induced processes.} 
	\label{fig:nloafb}%
\end{figure}

In Figs.~\ref{fig:nlomll}-\ref{fig:nloafb} we present the breakdown of the possible $\gamma$-induced contributions at fixed-order for the invariant-mass and the asymmetry distributions. 
For the $M(\mu^+\mu^-)$ distribution, we take ratios of predictions with and without $\gamma$-initiated processes, 
while for $A_{FB}$ we take absolute differences of the two types of predictions 
since $A_{FB}$ crosses zero around 90~GeV. 
In the green, red, and blue lines, both the contributions with and without $\gamma$-induced processes are computed using the {\tt NNPDF31\_nlo\_as\_0118\_luxqed} PDF set, 
as we want to single-out the impact of the $\gamma$-initiated contributions to the hard scattering process.
However, it is not entirely consistent to use the {\tt NNPDF31\_nlo\_as\_0118\_luxqed} PDFs in a calculation involving the $q\bar{q}$ channel only, 
as photons and in particular the $\gamma\to q\bar{q}$ splitting enter the predictions via PDFs evolution. 
For this reason, in Figs.~\ref{fig:nlomll}-\ref{fig:nloafb}, we added the solid black line where the predictions with $\gamma$-induced processes are computed using {\tt NNPDF31\_nlo\_as\_0118\_luxqed}, while the reference results without $\gamma$-initiated terms are based on the {\tt NNPDF31\_nlo\_as\_0118} PDF set.
Figure~\ref{fig:nlomll} shows that the impact of the NLO QED corrections to the $\gamma\gamma$ channel is very small and hardly visible on the scale of the plot, 
though it increases when moving away from the $Z$ peak. 
The $q\gamma$ and $\gamma\gamma$  contributions largely compensate each other below the resonance region, leading to $\mathcal{O}(1\%)$ effects in the low $M(\mu^+\mu^-)$ region, while above the resonance the $\gamma\gamma$ channel dominates over the $q\gamma$ one and the corresponding effect is of order 2\% at 150~GeV. 
If we do not include the $\gamma$ contributions to PDF evolution in the calculation of the reference predictions without $\gamma$-induced processes, the blue line is shifted downward by about $0.5\%$, thus reducing the impact of $\gamma$-induced processes off peak, but introducing a negative $0.5\%$ normalization effect at the $Z$ peak.
As far as the green, red, and blue lines are concerned, the picture for $A_{FB}$ is very similar to the one described for $M(\mu^+\mu^-)$, though in this case the $\gamma\gamma$ channel only contributes to the denominator of Eq.~(\ref{eq:afb}). 
Considering also the impact of photon splittings in PDF evolution, 
the correction from $\gamma$-induced processed is in the range of $0.01\%$ up to about 120~GeV and is below $10^{-5}$ at the $Z$ peak. 

\begin{figure}
	\includegraphics[width=0.47\textwidth]{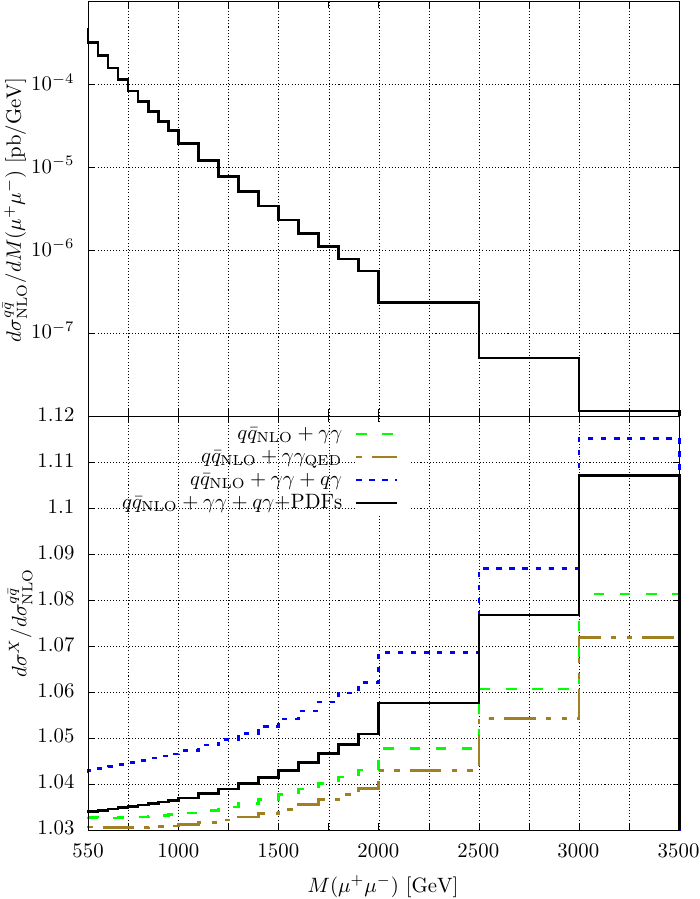} 
	\caption{Photon-induced corrections to the di-lepton invariant-mass in the TeV range. 
    Same notation and conventions as in Fig.~\ref{fig:nlomll}.} 
	\label{fig:nlomllhm}%
\end{figure}

\begin{figure}
	\includegraphics[width=0.47\textwidth]{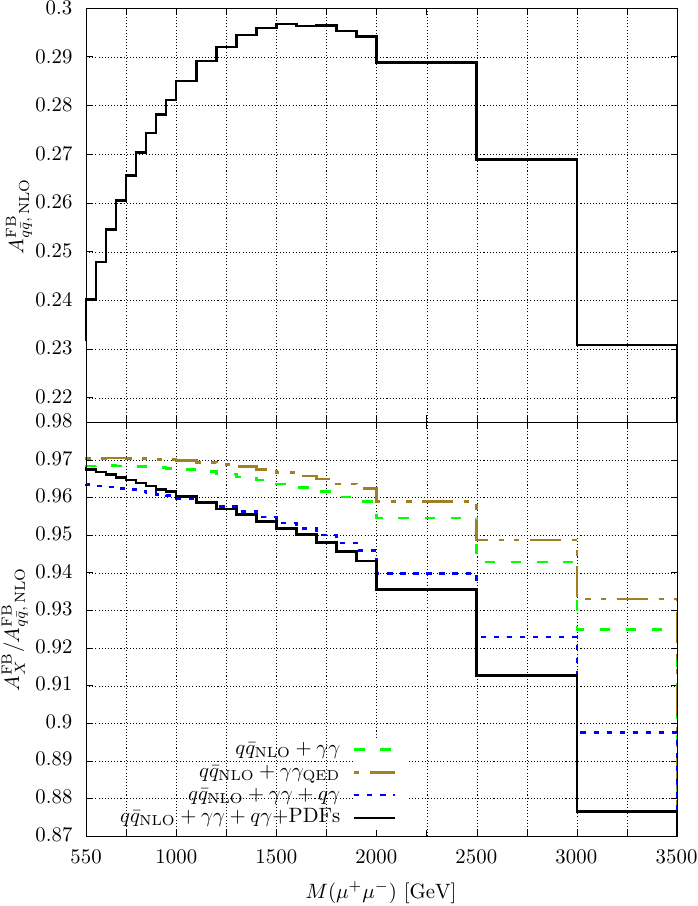} 
	\caption{Photon-initiated contributions to the forward-backward asymmetry in the TeV range. 
    Same notation and conventions as in Fig.~\ref{fig:nlomll}. 
    At variance with Fig.~\ref{fig:nloafb}, results are presented as ratios of predictions with and without $\gamma$-induced processes.} 
	\label{fig:nloafbhm}%
\end{figure}

Figures~\ref{fig:nlomllhm}-\ref{fig:nloafbhm} show the impact of $\gamma$-initiated processes on $M(\mu^+\mu^-)$ and $A_{FB}$ at high invariant-masses. 
As the asymmetry is positive in this region, we take ratios of predictions with/without $\gamma$-induced processes for both the observables.
The leading contribution to $M(\mu^+\mu^-)$ comes from the $\gamma\gamma$ UB and ranges from 3\% to 8\% at 3~TeV. 
QED corrections to the $\gamma\gamma$ channel tend to reduce the impact of the $\gamma\gamma$ process and reach $\mathcal{O}(1\%)$ in the right tail of the plot. 
The $q\gamma$ contributions are positive and go from about 1\% at 0.5~TeV up to 4\% at 3~TeV. 
When using the {\tt NNPDF31\_nlo\_as\_0118} PDF set for the predictions without $\gamma$-initiated process, the blue line is shifted by about $-1\%$. 
$\gamma$-induced contributions to $A_{FB}$ are negative and dominated by the corrections to the denominator of Eq.~(\ref{eq:afb}). In fact, if one considers only the impact on the hard scattering process, the relative effect of $\gamma$-induced corrections on $A_{FB}$ is basically equal and opposite to the one on $M(\mu^+\mu^-)$. 
On the other hand, the $\gamma$ contribution to PDFs affects in a  slightly different way the numerator and the  denominator of Eq.~(\ref{eq:afb}). As a result, the black line in Fig.~\ref{fig:nloafb} crosses the blue one around 1~TeV and the overall correction gets larger in absolute value, becoming of order $-12\%$ at 3~TeV.

\begin{figure}
    \includegraphics[width=0.47\textwidth]{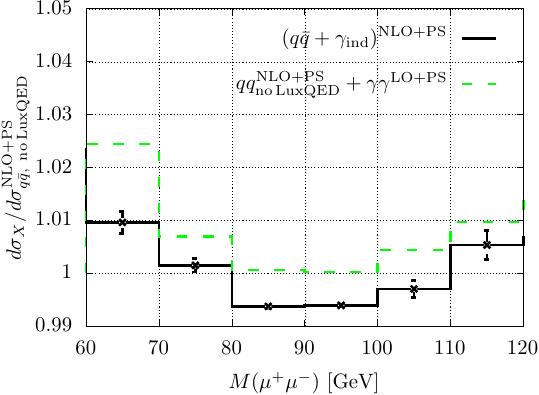} 
	\caption{Relative effect on the invariant-mass distribution of the $\gamma$-induced contributions at NLO QCD+NLO EW plus fermionic higher-orders matched to QED and QCD. The reference predictions in the denominator include only the $q\bar{q}$ channel computed using the {\tt NNPDF31\_nlo\_as\_0118} PDF set. 
    The numerator in the black line contains the NLO+PS predictions for the $q\bar{q}$ channel with the $q\gamma$ real contributions and the corresponding $\gamma\gamma$ UB. In the green dashed line, the numerator includes, besides the $q\bar{q}$ channel at NLO+PS, the $\gamma\gamma$ channel at LO+PS accuracy. 
    The {\tt NNPDF31\_nlo\_as\_0118\_luxqed} PDF is always used in the numerators.
    } 
	\label{fig:pymll}%
\end{figure}

\begin{figure}
    \includegraphics[width=0.47\textwidth]{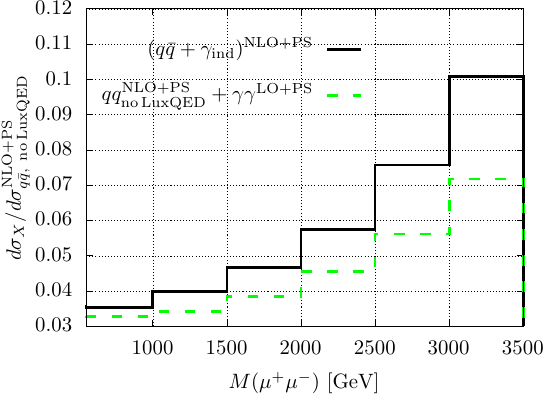} 
	\caption{Relative impact on the invariant-mass distribution of the $\gamma$-induced contributions at NLO+PS level. Same conventions as in Fig.~\ref{fig:pymll}.} 
	\label{fig:pymllHM}%
\end{figure}

\begin{figure}
    \includegraphics[width=0.47\textwidth]{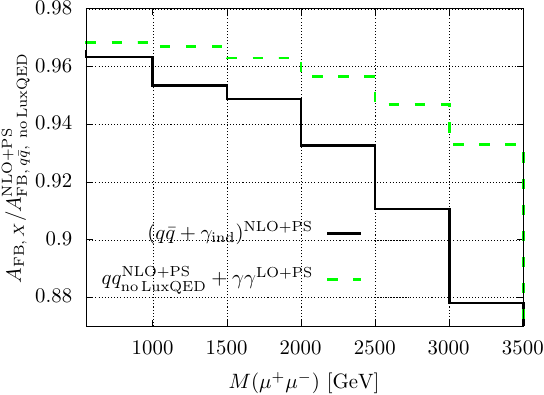} 
	\caption{Relative impact on $A_{\rm FB}$ of the $\gamma$-initiated processes at NLO+PS. Same conventions as in Fig.~\ref{fig:pymll}.} 
	\label{fig:pyafbHM}%
\end{figure}

In the following, we focus on the predictions at NLO QCD+NLO EW accuracy matched to QCD and QED PS. 
Figure~\ref{fig:pymll} shows, for the dilepton invariant-mass distribution, the relative difference between the results obtained with or without the $\gamma$-induced contributions.
Though the $M(\mu^+\mu^-)$ distribution receives corrections when going from NLO to NLO+PS, such effects are basically the same for both the calculations. 
As a result, the ratio in Fig.~\ref{fig:pymll} is close to the one observed at fixed-order in Fig.~\ref{fig:nlomll}. 
The plot also shows the $\gamma\gamma$ channel compared to the $q\bar{q}$ one when the former is computed at LO+PS (the PS starting scale being the partonic center-of-mass energy). 
Figure~\ref{fig:pymll} shows that estimating the $\gamma$-induced corrections using LO+PS predictions for $\gamma\gamma\to l^+l^-$ might lead to an overestimate of the effect at low invariant masses and for $M(\mu^+\mu^-)$ values in the 100~GeV range and to an underestimate of the correction at the $Z$ peak. 
Although  the dashed green curve in Fig.~\ref{fig:pymll} is similar to the one in Fig.~\ref{fig:nlomll}, there is no one-to-one correspondence since the PDF used in the denominators of the two curves are different.  
We refrain from showing results at NLO+PS for $A_{FB}$ since we found that the $\gamma$-induced contributions at fixed-order lead to corrections of order $10^{-4}$ or even smaller in the resonance region and huge event samples would be needed to be sensitive to such tiny effects. 

Figures~\ref{fig:pymllHM} and~\ref{fig:pyafbHM} replicate the analysis in Fig.~\ref{fig:pymll}, but in the high invariant-mass region. 
Results are shown for $M(\mu^+\mu^-)$ and $A_{FB}$. 
As observed before for the resonance region, the impact of the $\gamma$-induced processes at NLO+PS is close to the NLO one in Figs.~\ref{fig:nlomllhm}-\ref{fig:nloafbhm} (solid black line). 
In the high $M(\mu^+\mu^-)$ region, LO+PS simulations of the $\gamma\gamma$ channel undershoot the total $\gamma$-initiated contribution by an amount ranging from fractions of percent at 500~GeV up to order $5\% $ at 3~TeV for both observables.

\section{Summary and conclusions}
\label{sec:conclusions}

The precision physics program at the LHC and future colliders requires high-precision theoretical predictions not only for signal but also for background processes. 
For NC DY, $\gamma\gamma\to l^+l^-$ represents an irreducible background that, despite being strongly suppressed by  the photon PDF, remains relevant in high precision measurements at the weak scale and  can become sizable at  invariant masses in the TeV region. 
Furthermore, when considering predictions for NC DY beyond LO in the EW theory, the separation between the $q\bar{q}$ and the $\gamma\gamma$ channels is no longer straightforward. 

In this paper we presented an implementation of the $\gamma$-induced corrections to NC DY at NLO+PS accuracy in the POWHEG framework. We discussed the required modification to the {\tt POWHEG-BOX} routines, as well as the resulting predictions for the dilepton invariant mass and forward-backward asymmetry both in the $Z$ resonance region and at high invariant masses. 

We found that the part of the NLO+PS corrections beyond NLO are small compared to the scales considered in our analysis, 
but we also presented the shortfalls of estimating the $\gamma$-initiated contributions at LO+PS. 
The above consideration as well as the interplay between $q\bar{q}$ and $\gamma\gamma$ channels beyond LO motivate the development of Monte Carlo generators like {\tt Z\_ew-BMNNPV} including the $\gamma$-induced contributions at NLO+PS. More realistic phenomenological studies on the impact of the $\gamma$-induced corrections on the extraction of the effective or on the $\overline{\rm MS}$ weak mixing angle are left for future work.

\section*{Acknowledgements}
F.B.~and M.C.~would like to thank the kind hospitality of
the GGI Institute in Florence, Italy, during part of the work. C.D.P. is supported
by the U.S. Department of Energy under Contract No. DE-
SC0012704.
This work has been partially supported by the Italian Ministry of University
and Research (MUR), with EU funds (NextGenerationEU), through the PRIN2022
grant agreement Nr. 20229KEFAM (CUP I53D23001000006).

\bibliographystyle{elsarticle-num} 
\bibliography{mainbib}

\end{document}